\documentclass[]{rQUF2e}
\bibpunct{(}{)}{,}{a}{}{,}
\begin{document}
\doi{10.1080/1469768YYxxxxxxxx}
 \issn{1469-7696} \issnp{1469-7688} \jvol{00} \jnum{00} \jyear{2008} \jmonth{July}

\markboth{Linlin Xu and Giray \"{O}kten}{High Performance Financial Simulation Using Randomized Quasi-Monte Carlo Methods}

\title{High Performance Financial Simulation Using Randomized Quasi-Monte Carlo Methods}

\author{Linlin Xu$\dag$ and Giray \"{O}kten$^{\ast}$$\dag$\thanks{$^\ast$Corresponding author. Email: okten@math.fsu.edu
\vspace{12pt}}\\\vspace{12pt}  
\normalfont{$\dag$Department of Mathematics, Florida State Univesity, Tallahassee, FL 32306
}\\\vspace{12pt}\received{v1.2 released October 2008} }

\maketitle

\begin{abstract}
GPU computing has become popular in computational finance and many
financial institutions are moving their CPU based applications to the GPU
platform. Since most Monte Carlo algorithms are embarrassingly parallel, they
benefit greatly from parallel implementations, and consequently Monte Carlo
has become a focal point in GPU computing. GPU speed-up examples reported in
the literature often involve Monte Carlo algorithms, and there are software
tools commercially available that help migrate Monte Carlo financial pricing models to GPU.

We present a survey of Monte Carlo and randomized quasi-Monte Carlo
methods, and discuss existing (quasi) Monte Carlo sequences in GPU libraries.
We discuss specific features of GPU architecture relevant for developing
efficient (quasi) Monte Carlo methods. We introduce a recent randomized quasi-Monte Carlo method, and compare it with some of the existing
implementations on GPU, when they are used in pricing caplets in the LIBOR market model and mortgage backed securities.

\newpage

\end{abstract}
\vspace{-50pt}
\section*{}

\section{Introduction}

The recent trend towards parallel computing in the financial industry is not surprising. As the complexity of models used in the industry grows, while the demand for fast, sometimes real-time, solutions persists, parallel computing is a resource that is hard to ignore. In 2009, Bloomberg and NVIDIA worked together to run a two-factor model for calculating hard-to-price asset-backed securities on 48 Linux servers paired with Graphics Processing Units (GPUs), which traditionally required about 1000 servers to accommodate customer demand. GPU computing offers several advantages over traditional parallel computing on clusters of CPUs. Clusters consume non negligible energy and space, and computations over clusters are not always easy to scale. In contrast, GPU is small, fast, and consumes only a tiny fraction of energy consumed by clusters. Consequently, there has been a recent surge in academic papers and industry reports that document benefits of GPU computing in financial problems. Arguably, the numerical method that benefits most from GPUs is the Monte Carlo simulation. Monte Carlo methods are inherently parallel, and thus more suitable for implementing on GPU than most alternative methods. In this paper we concentrate on Monte Carlo methods and financial simulation, and discuss computational and algorithmic issues when financial simulation algorithms are developed over GPU and traditional clusters.

The computational framework we use is the estimation of an
integral $I=\int_{(0,1)^{s}}f(x)dx$ over the $s$ dimensional unit cube, using
sums of the form $\theta_{N}=\frac{1}{N}\sum_{i=1}^{N}f(x_{i}).$ In Monte
Carlo and quasi-Monte Carlo, $\theta_{N}$ converges to $I$ as $N\rightarrow
\infty.$ In the former the convergence is probabilistic and $x_{i}$ come from
a pseudorandom sequence, and in the latter the convergence is deterministic
and the $x_{i}$ come from a low-discrepancy sequence. For a comprehensive
survey of Monte Carlo and quasi-Monte Carlo methods, see \cite{Nied}. Often it is desirable to obtain multiple independent estimates
for $\theta,$ say $\theta^{1},...,\theta^{m},$ so that one could use
statistics to measure the accuracy of the estimation by the use of sample
standard deviation, or confidence intervals. Let us assume that an allocation of
computing resources is done and we choose parameters $N,M$: the first
parameter, $N,$ is the sample size, and gives the number of vectors from the
sequence (pseudorandom or low discrepancy) to use in estimating $\theta
:=\theta_{N}^{m}$
\[
\theta_{N}^{m}=\frac{1}{N}\sum_{i=1}^{N}f(q_{i}^{m})
\]
and the parameter $M$ gives the number of independent replications we obtain
for $\theta_{N}$, i.e., $\theta_{N}^{1},...,\theta_{N}^{M}.$ The grand average
$RQ_{M,N}$ gives the overall point estimate for $I:$%
\[
RQ_{M,N}=\frac{1}{M}\sum_{m=1}^{M}\theta_{N}^{m}.
\]
In Monte Carlo, to obtain the independent estimates $\theta_{N}^{1}%
,...,\theta_{N}^{M},$ one simply uses blocks of $N$ pseudorandom numbers. In
quasi-Monte Carlo, one has to use methods that enable independent
randomizations of the underlying low-discrepancy sequence. These methods are
called randomized quasi-Monte Carlo (RQMC) methods (see \citet{okten warren, okten}).

\begin{figure}
\begin{center}
\includegraphics[
height=1.5in,
width=4.5in
]
{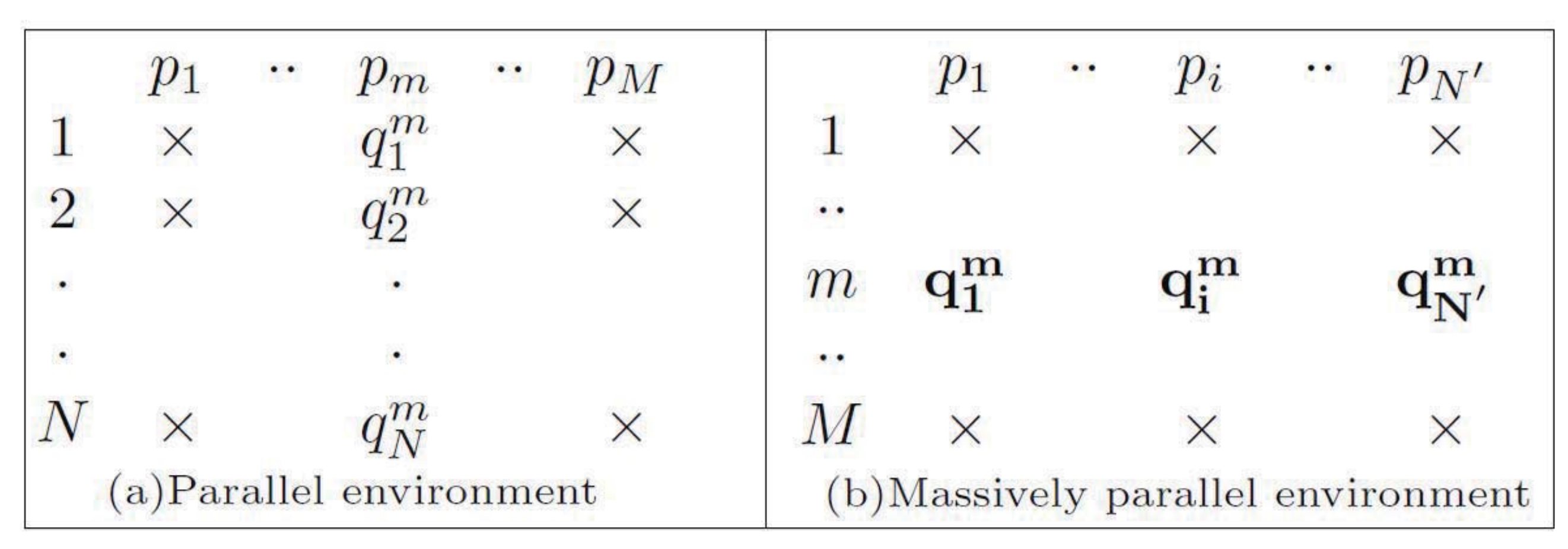}
\end{center}
\caption{Different parallel computing strategies}%
\label{paral}
\end{figure}

Traditionally, in parallel implementations of Monte Carlo algorithms, one
often assigns the $m$th processor (of the $M$ allocated processors) the
evaluation of the estimate $\theta_{N}^{m}.$ To do this computation, each
processor needs to have an assigned number sequence (pseudorandom or
low-discrepancy) and methods like blocking, leap-frogging, and
parameterization are used to make this assignment. Parameterization is
particularly useful when independent replications are needed to compute
$RQ_{M,N}$ (see \citet{okten matt}, and also \citet{DZCG}, \citet{Zinterhof seq},
\citet{okten ashok}). If only a single estimate is needed, then blocking or
leap-frogging can be used (\citet{bromley}, \citet{chen}, \citet{LM},
\citet{schmiduhl1}, \citet{schmiduhl2}). Figure \ref{paral}(a) describes this traditional Monte Carlo implementation where the $m$th processor $p_{m}$ generates its assigned sequence $q_{1}^{m},...,q_{N}^{m}$ to compute $\theta_{N}^{m},$ as $m=1,...,M$. In many applications $N$ is typically in millions, and $M$ is large enough for statistical accuracy, in the range 50 to 100. 

In a massively parallel environment, depicted by the second diagram, where the number of processors $N^{'}$is much larger than $M$, it can be a lot more efficient to completely
``transpose'' our computing strategy. Now the processors $p_{1},...,p_{N^{'}}$ run
simultaneously (for a total of $M$ times) to generate the sequence $\mathbf{q_{1}^{m}},\cdots,\mathbf{q_{N^{'}}^{m}}$ to compute $\theta_{N}^{m},$ as $m=1,...,M$, where $\mathbf{q_{i}^{m}}$ is part of the sequence $\{q_{i}^{m},q_{N^{'}+i}^{m}, q_{2N^{'}+i}^{m},... \}$ which is assigned to the $i$th processor.

The choice of the two computing paradigms, which we vaguely name as ``parallel''
and ``massively parallel'', determines how the underlying sequence (pseudorandom
or low-discrepancy) should be generated. In the parallel paradigm, a recursive
algorithm for generating the underlying sequence works best since each
processor generates the ``entire'' sequence. This paradigm is appropriate for a computing system with distributed memory, such as a cluster. For the massively parallel
paradigm, a direct algorithm that generates the $n$th term of the sequence from $n$ is more appropriate. \cite{shaw} use the term ``counter-based'' to describe such direct algorithms. The massively parallel paradigm is an appropriate model for GPU computing where prohibitive cost of memory access makes recursive computing inefficient.

In Section \ref{Section pseudorandom} we briefly discuss a counter-based pseudorandom number generator, called \textbf{Philox}, introduced by \cite{shaw}, and
the pseudorandom number generators, \textbf{Mersenne
twister}, and \textbf{XORWOW}. In Section \ref{Section rstart} we introduce a randomized
quasi-Monte Carlo sequence, which we name \textbf{Rasrap}, and give algorithms
for recursive and counter-based implementations of this sequence. In this section, we also give a brief description of a well-known quasi-Monte Carlo sequence, the Sobol' sequence. We will
compare the computational time for generating these sequences on CPU and GPU,
in Section \ref{Section timing comp}.
\section{Monte Carlo sequences\label{Section pseudorandom}}

Most pseudorandom number generators are inherently iterative: they are
generated by successive application of a transformation $F$ to an element of
the state space to obtain the next element of the state space, i.e.,
$s_{n}=F(s_{n-1})$. Here we discuss some of the pseudorandom number generators considered in this paper. One of the most popular and high quality pseudorandom number generators is the Mersenne twister introduced by \citet{twister}. It has a very
large period and excellent uniformity properties. It is available in many
platforms, and recently Matlab adopted it as its default random number generator.

A parallel implementation of the Mersenne twister was also given by 
\citet{parallel twister}. Their approach uses parameterization, and it falls
under our parallel computing paradigm: each processor in the parallel
environment generates a Mersenne twister, and different Mersenne twisters
generated across different processors are assumed to be statistically
independent. There are several parameters that need to be precomputed and
stored to run the parallel implementation of Mersenne twister.

XORWOW is a fast pseudorandom number generator introduced by \citet{xor}. This generator is available in CURAND: a library for pseudorandom and quasi-random number generators for GPU provided by NVIDIA. However, the generator fails certain statistical tests; see \citet{saito} for a discussion. The reason we consider this generator is because of its availability in CURAND, and that its computational speed can be used as a benchmark against which other generators can be compared.

Philox is a counter-based pseudorandom number generator introduced by \citet{shaw}. Its generation is in the form $s_{n}=F(n)$, and thus
falls under our massively parallel computing paradigm. A comparison of some
counter-based and conventional pseudorandom number generators (including
Philox and Mersenne twister) is given in \citet{shaw}. In Section
\ref{Section timing comp}, we will present timing results comparing the pseudorandom number generators, and in Section \ref{LIBOR} and \ref{MBS}, we will compare
these sequences when they are used in some financial problems. These numerical
results will also include Rasrap and Sobol', two randomized-quasi Monte Carlo sequences that we discuss next.

\section{Randomized-quasi Monte Carlo sequences\label{Section rstart}}
\subsection{Rasrap}
The van der Corput sequence, and its generalization to higher dimensions, the
Halton sequence, are among the best well-known low-discrepancy sequences. The
$n $th term of the van der Corput sequence in base $b $, $\phi_{b}(n) $, is
defined as

\begin{equation}
\label{vdcp}\phi_{b}(n) = (.a_{0} a_{1} \cdots a_{k})_{b} = \frac{a_{0}}{b} +
\frac{a_{1}}{b^{2}} + \cdots+ \frac{a_{k}}{b^{k+1}},
\end{equation}
where

\begin{equation}
n=(a_{k}\cdots a_{1}a_{0})_{b}=a_{0}+a_{1}b+\cdots+a_{k}b^{k}. \label{expan}%
\end{equation}
The Halton sequence in the bases $b_{1},\cdots,b_{s}$ is $(\phi_{b_{1}%
}(n),\cdots,\phi_{b_{s}}(n))_{n=0}^{\infty}$. This is a low-discrepancy
sequence if the bases are relatively prime. In practice, $b_{i}$ is usually
chosen as the $i$th prime number.

There is a well-known defect of the Halton sequence: in higher dimensions,
when the base is larger, certain components of the sequence exhibit very poor
uniformity. This is often referred to as \textit{high correlation between large bases.}
As a remedy, permuted (or, scrambled) Halton sequences were introduced. The
\textit{permuted van der Corput sequence} generalizes (\ref{vdcp}) as
\begin{equation}
\phi_{b}(n)=\frac{\sigma(a_{0})}{b}+\frac{\sigma(a_{1})}{b^{2}}+\cdots
+\frac{\sigma(a_{k})}{b^{k+1}}, \label{perm}%
\end{equation}
where $\sigma$ is a permutation on the digit set $\{0,\cdots,b-1\}$. By using
different permutations for each base, one can define the permuted Halton
sequences in the usual way. There are many choices for permutations published
in the literature; a recent survey is given by \citet{Cools}. In this paper, we will follow the approach used in \citet{okten shah goncharov} and pick these permutations at random.

The Halton sequence can be generated recursively, which would be appropriate
for an implementation on CPU, or directly (counter-based), which would be
appropriate for GPU. Next we discuss some recursive and counter-based
algorithms for the Halton sequence.

A fast recursive method for generating the van der Corput sequence was given
by \citet{struckmeier}. We now explain his algorithm. Let $p$ be a
positive integer and $x\in\lbrack0,1)$ arbitrary. Define the sequence
$(b_{k}^{p})_{k\in\mathbb{N}}$ by
\begin{equation}
b_{k}^{p}=\frac{1}{p^{k}}(p+1-p^{k})\;\forall k\in\mathbb{N}, \label{bk}%
\end{equation}
and the transformation $T_{p}$ by
\begin{equation}
T_{p}(x)=x+b_{k}^{p}, \label{tp}%
\end{equation}
where\
\begin{equation}
k=\left\lfloor -\frac{\ln(1-x)}{\ln p}\right\rfloor +1. \label{k}%
\end{equation}
The transformation $T_{p}$ is called the von Neumann - Kakutani transformation
in base $p.$ The orbit of zero under $T_{p},$ i.e., $\{0,T_{p}(0),T_{p}%
^{2}(0),...\}$ is the van der Corput sequence in base $p$. In fact, the orbit
of any point $x_{0}\in\lbrack0,1)$ under $T_{p}$ is a low-discrepancy
sequence. If $x_{0}$ is chosen at random from the uniform distribution on
$[0,1),$ then the orbit of $x_{0}$ under $T_{p}$ is called a \textit{random-start van
der Corput sequence} in base $p.$ The following algorithm summarizes the
construction by \citet{struckmeier} of the (random-start) van der
Corput sequence in base $p.$ It can be generalized to Halton sequences in the
obvious way.

\begin{algorithm}
\citet{struckmeier}\label{struck alg}. Generates a random-start van
der Corput sequence with starting point $x_{0}$ and base $p.$

\begin{enumerate}
\item[(1)] Generate the sequence $b_{k}^{p}$ according to (\ref{bk});

\item[(2)] Choose an arbitrary starting point $x\in\lbrack0,1)$;

\item[(3)] Calculate $k$ according to (\ref{k});

\item[(4)] $x = x+b_{k}^{p}$;

\item[(5)] Repeat step 3-4.
\end{enumerate}
\end{algorithm}

Algorithm \ref{struck alg} is prone to rounding error in floating number
operations due to the floor operation in (\ref{k}). For example, a C++
compiler gives a wrong index $k$ after 3 steps of iteration when the starting
point is $x=0$ if the rounding error introduced in (\ref{bk}) is not carefully handled.

We now suggest an alternative algorithm that computes a random-start permuted
Halton sequence. The advantages of this algorithm over Algorithm
\ref{struck alg} are: (i) it avoids rounding errors, (ii) it is faster, and
(iii) it can be used to generate permuted Halton sequences.

\begin{algorithm}\label{recursive perm vdc}
(Recursive) Generates a random-start permuted van der Corput sequence in base $p$.

\begin{enumerate}
\item[(1)] Initialization Step. Generate a random number $\omega\in\lbrack0,1)$ and
find some integer $n$ so that $\omega$ is the $n^{th}$ term in the van Corput
sequence in base $p$. Initialize and store a random digit permutation $\sigma
$. Expand $n$ in base $p$ as $n=(a_{k}\cdots a_{1}a_{0})_{p}$ ($k$ depends on
$n$). Set $a_{i}=0$ for $i>k$. Store $a_{k},\cdots,a_{1},a_{0}$. Calculate and
store $S_{j}=\sum_{i=j}^{k}\frac{\sigma(a_{i})}{p^{i+1}}$ for $j=k,k-1,\cdots
,1,0$. Set $S_{j}=0$ for $j>k$. Set the quasi-random number $r=S_{0}$;

\item[(2)] Let $n=n+1$. Find $\min\{m|a_{m}+1<p\}$;

\item[(3)] $S_{m}=S_{m+1}+\frac{\sigma(a_{m}+1)}{p^{m+1}}$. Set $a_{m}=a_{m}+1$.
Set $a_{i}=0,S_{i}=S_{i+1}+\frac{\sigma(0)}{p^{i+1}}$ for $i=m-1,m-2,\cdots
,1,0 $. The quasi-random number corresponding to $n+1$ is $r=S_{0}$;

\item[(4)] Repeat step 2-3.
\end{enumerate}
\end{algorithm}

Algorithm 2 is an efficient iterative algorithm appropriate for the parallel
computing paradigm. However, for the massively parallel computing paradigm,
such as GPU computing, we need a counter-based algorithm. For the Halton
sequence, this would be simply its definition:

\begin{algorithm}
(Counter-based) \label{definition vdc} Generates a random-start permuted van der
Corput sequence in base $p$.

\begin{enumerate}
\item[(1)] Initialization step: Choose a small positive real number, $\epsilon$. Generate a random number $\omega$ from the uniform distribution on $(0,1)$, and
find $n$ such that $|\phi_{p}(n)-\omega|<\epsilon$;

\item[(2)] The quasi-random number corresponding to $n$ is $\phi_{p}(n)$;

\item[(3)] Let $n=n+1$ and repeat step 2-3.
\end{enumerate}
\end{algorithm}

The name Rasrap is an abbreviation for random-start randomly permuted Halton
sequence: if in Algorithms \ref{recursive perm vdc} and \ref{definition vdc},
the permutations for each base are generated at random, then we obtain Rasrap.

\subsection{Sobol' sequence}
The Sobol' sequence is a well-known fast low-discrepancy sequence popular among financial engineers.
The $j$th component of the $i$th vector in a Sobol' sequence is calculated by
\begin{center}
$x_{i}^j=i_{1}v_{1}^j\oplus i_{2}v_{2}^j\oplus\cdots$,
\end{center}
where $i_{k}$ is the $k$th digit from the right when integer $i$ is represented in base $2$ and $\oplus$ is the bitwise exclusive-or operator. The so-called direction numbers, $v_{k}^j$, are defined as

\begin{center}
$v_{k}^j = \frac{m_{k}^j}{2^{k}}.$
\end{center}
To generate the Sobol' sequence, we need to generate a sequence of positive integers $\{m_{k}^j\}$. The sequence $\{m_{k}^j\}$ is defined recursively as follows:
\begin{center}
$m_{k}^j = 2a_{1}^j m_{k-1}^j \oplus 2^{2}a_{2}^j m_{k-2}^j \oplus \cdots \oplus 2^{s_{j}-1}a_{s_{j}-1}^j m_{k-s_{j}+1}^j\oplus 2^{s_{j}}m_{k-s_{j}}^j \oplus  m_{k-s_{j}}^j $,
\end{center}
where $a_{1}^j, a_{2}^j, \cdots, a_{s_{j}-1}^j$ are coefficients of a primitive polynomial of degree $s_{j}$ in the field $\mathbb{Z}_{2}$,
\begin{center}
$x^{s_{j}} + a_1^jx^{s_j-1} + a_2^jx^{s_j-2} + \cdots + a_{s_j-1}^jx + 1$.
\end{center}

The initial values $m_1^j, m_2^j, \cdots, m_{s_j}^j$ can be chosen freely given that each $m_l^j, 1 \leq l \leq s_j$, is odd and less than $2^l$. Because of this freedom, different choices for direction numbers can be made based on different search criteria minimizing the discrepancy of the sequence. We use the primitive polynomials and direction numbers provided by \citet{js08}.

The counter-based implementation of the Sobol' sequence introduced here is convenient on GPUs, but a more efficient implementation proposed by Antonov and Saleev based on Gray code is used in practice on CPUs. For details about this approach, see \citet{as79}.

The Sobol' sequence can be randomized using various randomized quasi-Monte Carlo methods. Here we will use the random digit scrambling method of \cite{matousek}. More on randomized quasi-Monte Carlo and some parallel implementations can be found in \cite{okten warren}, and, \cite{okten matt}.

\section{Performance Comparison\label{Section timing comp}}

Mersenne twister, Philox, XORWOW, Rasrap, and Sobol' sequences are run on Intel i7 3770K and NVIDIA GeForce GTX 670. We compare the throughput of different algorithms on CPU (Table 1) and GPU (Table 2).

Table 1 shows that the fastest algorithm for the Halton sequence on CPU is
Algorithm 2. It is about 3.8 times as fast as the algorithm by Struckmeier
(Algorithm 1). Not surprisingly Algorithm 3, the counter-based implementation,
is considerably slower on CPU. Mersenne twister uses its serial CPU
implementation and it is about 3.4 times faster than Algorithm 2 for the
Halton sequence. And Sobol' sequence based on Gray code is faster than Mersenne twister.

Table 2 shows that the throughput of Algorithm 3 on GPU improves significantly compared to the CPU value. Counter-based Sobol' sequence is twice as fast as Rasrap, and the pseudorandom number generator Philox is almost 200 times faster than Rasrap.

\begin{table}\label{genCPU}
\begin{center}
\begin{minipage}{80mm}
  \tbl{Throughput of generators on CPU.} 
{\begin{tabular}{@{}lcc}\toprule
   
  & Throughput (GNumbers/s) \\
\colrule
	Twister & 0.598\\
	Rasrap Algo. \ref{struck alg} & 0.045\\
	Rasrap Algo. \ref{recursive perm vdc} & 0.173\\
	Rasrap Algo. \ref{definition vdc} & 0.012\\
	Sobol'(Counter based) & 0.04\\
	Sobol'(Gray Code) & 0.97\\
  \botrule
\end{tabular}}
\end{minipage}
\end{center}
\end{table}

\begin{table}
\begin{center}
\begin{minipage}{80mm}
  \tbl{Throughput of generators on GPU.}
{\begin{tabular}{@{}lcc}\toprule
   
  & Throughput (GNumbers/s) \\
\colrule
	XORWOW & 60\\
	Philox & 190\\
	Rasrap Algo. \ref{definition vdc} & 1.0\\
	Sobol'(Counter based) & 2.0 \\
  \botrule
\end{tabular}}

\end{minipage}
\end{center}
\end{table}

The computational speed at which various sequences are generated is only one
part of the story. We next examine the accuracy of the estimates obtained when
these sequences are used in simulation. In the next section, we use these
sequences in two problems from computational finance, and compare them with
respect to the standard deviation of their estimates and computational speed.

\section{Pricing caplets in the LIBOR model}

\label{LIBOR}
An interest rate derivative is a derivative where the underlying asset is the right to pay or receive a notional amount of money at a given interest rate. The interest rate derivatives market is the largest derivatives market in the world. To price interest rate derivatives, forward interest rate models are widely used in the industry. There are two kinds of forward rate models: the continuous rate model and the simple rate model.

The framework developed by \citet{hjm92} (HJM) explicitly describes the dynamics of the term structure of the interest rates through the dynamics of the forward rate curve. HJM model has two major drawbacks: (1) the instantaneous forward rates are not directly observable in the market; (2) some simple choices of the form of volatility is not admissible. 

In practice, many fixed income securities quote the interest rate on an annual basis with semi-annual or quarterly compounding, instead of a continuously compounded rate. The simple forward rate models describe the dynamics of the term structure of interest rates through simple forward rates, which are observable in the market. This approach is developed by \citet{mss97}, \citet{bgm97}, \citet{mr97} and \citet{jf97}. 

The London Inter-Bank Offered Rates (LIBOR) is one of the most important benchmark simple interest rates. Let $ B(t, T) $ denote the time-$ t $ value of a zero coupon bond paying 1 at the maturity time $ T $. A forward rate $ F(t, T_1, T_2) $ ($ t < T_1 < T_2 $ ) is an interest rate fixed at time $ t $ for borrowing or lending at time $ T_1 $ over the period $ [T_1, T_2] $. An arbitrage argument shows that forward rates are determined by bond prices in accordance to
\begin{equation}
\label{fwdrate}
F(t, T_1, T_2) = \frac{1}{T_2 - T_1}\bigg ( \frac{B(t, T_1) - B(t, T_2)}{B(t, T_2)}\bigg ).
\end{equation}
A forward LIBOR rate is a special case of (\ref{fwdrate}) with a fixed period $ \delta = T_2 - T_1 $ for the accrual period. Typically $ \delta = 0.5$ or $0.25 $. Thus, the $ \delta $-year forward LIBOR rate at time $ t $ with maturity $ T $ is
\begin{equation}
\label{fwdLIBOR}
L(t, T) = F(t, T, T + \delta) = \frac{1}{\delta}\bigg ( \frac{B(t, T) - B(t, T + \delta)}{B(t, T + \delta)}\bigg ).
\end{equation}

So if we enter into a contract at time $ 0 $ to borrow 1 at time $ T $ and repay it with interest at time $ T + \delta $, the interest due will be $ \delta L(0, T) $. 

Fix a finite set of maturities
\begin{center}
$ 0 = T_0 < T_1 < \cdots < T_M < T_{M+1} $
\end{center}
and let 
\begin{center}
$ \delta_i = T_{i+1} - T_i,\;i = 0, \cdots, M $,
\end{center}
denote the lengths of the intervals between maturities. Normally we fix $ \delta $ as a constant regardless of day-count conventions that would introduce slightly different values for the fractions $ \delta_i $.

For each maturity $ T_n $, let $ B_n(t) $ denote the time-$ t $ value of a zero coupon bond maturing at $ T_n,\; 0\leq t\leq T_n $. And write $ L_n(t) $ for the forward rate at time $ t $ over the period $ [T_n, T_{n+1}] $. Equation (\ref{fwdLIBOR}) can be then rewritten as
\begin{equation}
\label{fwdLIBORn}
L_n(t) = \frac{B_n(t)- B_{n+1}(t)}{\delta_n B_{n+1}(t)},\; 0\leq t \leq T_n,\; n = 0, 1, \cdots, M.
\end{equation}
The subscript $ n = 0, 1, \cdots, M $ emphasizes we are looking at a finite set of bonds.

The dynamics of the forward LIBOR rates can be described as a system of SDEs as follows. For a brief informal derivation, see \citet{glasserman03}.
\begin{equation}
\label{dL_n}
\frac{dL_n(t)}{L_n(t)} = \sum_{j=\eta(t)}^n\frac{\delta_j(t) L_j(t)\sigma_n(t)^\top \sigma_j(t)}{1+\delta_j L_j(t)} dt + \sigma_n(t)^\top dW(t),\; 0\leq t \leq T_n,\; n = 1, \cdots, M.
\end{equation}
where $ W $ is a $ d $-dimensional standard Brownian motion and the volatility $ \sigma_n $ may depend on the current vector of rates $ (L_1(t),\cdots, L_M(t)) $ as well as the current time $ t $. $ \eta(t) $ is the unique integer such that $ T_{\eta(t)-1} \leq t < T_{\eta(t)}$.

Pricing interest rate derivative securities with LIBOR market models normally requires simulations. Since the LIBOR market model deals with a finite number of maturities, only the time variable needs to be discretized.

We fix a time grid $ 0 = t_0 < t_1 <\cdots < t_m < t_{m+1} $ to simulate the LIBOR market model. In practice, one would often take $ t_i = T_i $ so the simulation goes directly from one maturity date to the next. For simplicity, we use a constant volatility $ \sigma $ in the simulation. We apply an Euler scheme to (\ref{dL_n}) to discretize the system of SDEs of the LIBOR market model, producing
\begin{equation}
\label{L_n_discrete}
\hat{L}_n(t_{i+1}) = \hat{L}_n(t_i) + \mu_n(\hat{L}(t_i), t_i)\hat{L}_n(t_i)[t_{i+1} - t_i] + \hat{L}_n(t_i)\sqrt{t_{i+1} - t_i} \sigma_n(t_i)^{\top}Z_{i+1},
\end{equation}
where
\begin{equation}
\mu_n(\hat{L}(t_i), t_i) = \sum_{j=\eta(t_i)}^n \frac{\delta_j\hat{L}_j(t_i)\sigma_n(t_i)^{\top}\sigma_j(t_i)}{1 + \delta_j\hat{L}_j(t_i)}
\end{equation}
and $ Z_1, Z_2, \cdots $ are independent $ N(0, I)$ random vectors in $ \mathbb{R}^d $. Here hats are used to identify discretized variables. 

We assume an initial set of bond prices $ B_1(0), \cdots, B_{M+1}(0) $ is given and initialize the simulation by setting
\begin{equation}
\label{BL}
\hat{L}_n(0) = \frac{B_n(0) - B_{n+1}(0)}{\delta_nB_{n+1}(0)},\; n = 1, \cdots, M,
\end{equation}
in accordance with (\ref{fwdLIBORn}).

Next we use the simulated evolution of LIBOR market rates to price a caplet. An interest rate cap is a portfolio of options that serve to limit the interest paid on a floating rate liability over a set of consecutive periods. Each individual option in the cap applies to a single period and is called a caplet. It is sufficient to price caplets since the value of a cap is simply the sum of the values of its component caplets.

We follow the derivation in \citet{glasserman03}. Consider a caplet for the time period $ [T,T + \delta] $. A party with a floating rate liability over that period would pay interest $ \delta L(T, T) $ times the principle at time $ T+\delta $. A caplet is designed to limit the interest paid to a fixed level $ K $. The difference $ \delta(L(T, T) - K) $ would be refunded only if it is positive. So the payoff function of a caplet is
\begin{center}
$ \delta(L(T, T) - K)^{+} $,
\end{center}
where the notation $ (\cdot)^{+} $ indicates that we take the maximum of the expression in parentheses and zero.
This payoff is exercised at time $ T + \delta $ but determined at time $ T $. There is no uncertainty in the payoff over the period $ [T, T + \delta] $. Then the payoff function at time $ T + \delta $ is equal to
\begin{equation}
\frac{\delta(L(T, T) - K)^{+}}{1 + \delta L(T, T)} = \delta B(T, T + \delta)(L(T, T) - K)^{+}
\end{equation}
at time $ T $. This payoff typically requires the simulation of the dynamics of the term structure.

\citet{black76} derived a formula for the time-$ t $ price of the caplet under the assumption of $ L_n(T_n) $ following a lognormal distribution, which does not necessarily correspond to a price in the sense of the theory of derivatives valuation. In practice, this formula
\begin{eqnarray}\nonumber
\label{black_cap}
\sigma B(t, T + \delta)\Bigg(L(t, T)\Phi \bigg( \frac{\log(L(t, T)/K) + \sigma^2(T-t)/2}{\sigma \sqrt{T - t}} \bigg) \\
- K\Phi \bigg(\frac{\log(L(t, T)/K) - \sigma^2(T-t)/2}{\sigma \sqrt{T - t}} \bigg) \Bigg)
\end{eqnarray}
is used to calculate the ``implied volatility'' $ \sigma $ from the market price of caps.

To test the correctness of the LIBOR market model simulation, we use the daily treasury yield curve rates on 02/24/2012 as shown in Table 3 to initialize the LIBOR market rates simulation. We first apply a cubic spline interpolation to the rates in Table 3 to get estimated yield curve rates for every 6 months. Then the estimated yield curve rates are used to calculate the bond prices for every 6 months in order to initialize the LIBOR rates in (\ref{BL}). We assume the following parameters in LIBOR rates simulation
\begin{center}
$ (t, T, \delta, K, \sigma) = (0, 5, 0.5, 0.01, 0.04). $
\end{center} 

\begin{table}
\begin{center}
\begin{minipage}{80mm}
  \tbl{The daily Treasury yield curve rates on 02/24/2012.} 
{\begin{tabular}{@{}lcccccccccccc}\toprule
   
  Date & 1 mo & 3 mo & 6 mo & 1 yr & 2 yr & 3 yr& 5 yr & 7 yr & 10 yr & 20 yr & 30 yr \\
\colrule
	02/24/2012 & 0.08 & 0.10 & 0.14 & 0.18 & 0.31 & 0.43 & 0.89 & 1.41 & 1.98 & 2.75 & 3.10\\
  \botrule
\end{tabular}}
\end{minipage}
\end{center}
\end{table}
The simulations are run on Intel i7 3770K and NVIDIA GeForce GTX 670
respectively. For a fixed sample size $N$, we repeat the simulation 100 times using independent realizations of the underlying sequence. We investigate the sample standard deviation of the 100 estimates and computing time as a function of the sample size $N$. We also compare the efficiency of different sequences, where efficiency is defined as the product of sample standard deviation and execution time.

\subsection{Comparison of Sobol' sequence implementations}
\begin{figure}

\begin{center}
\includegraphics[
height=4.5in,
width=5.5in
]
{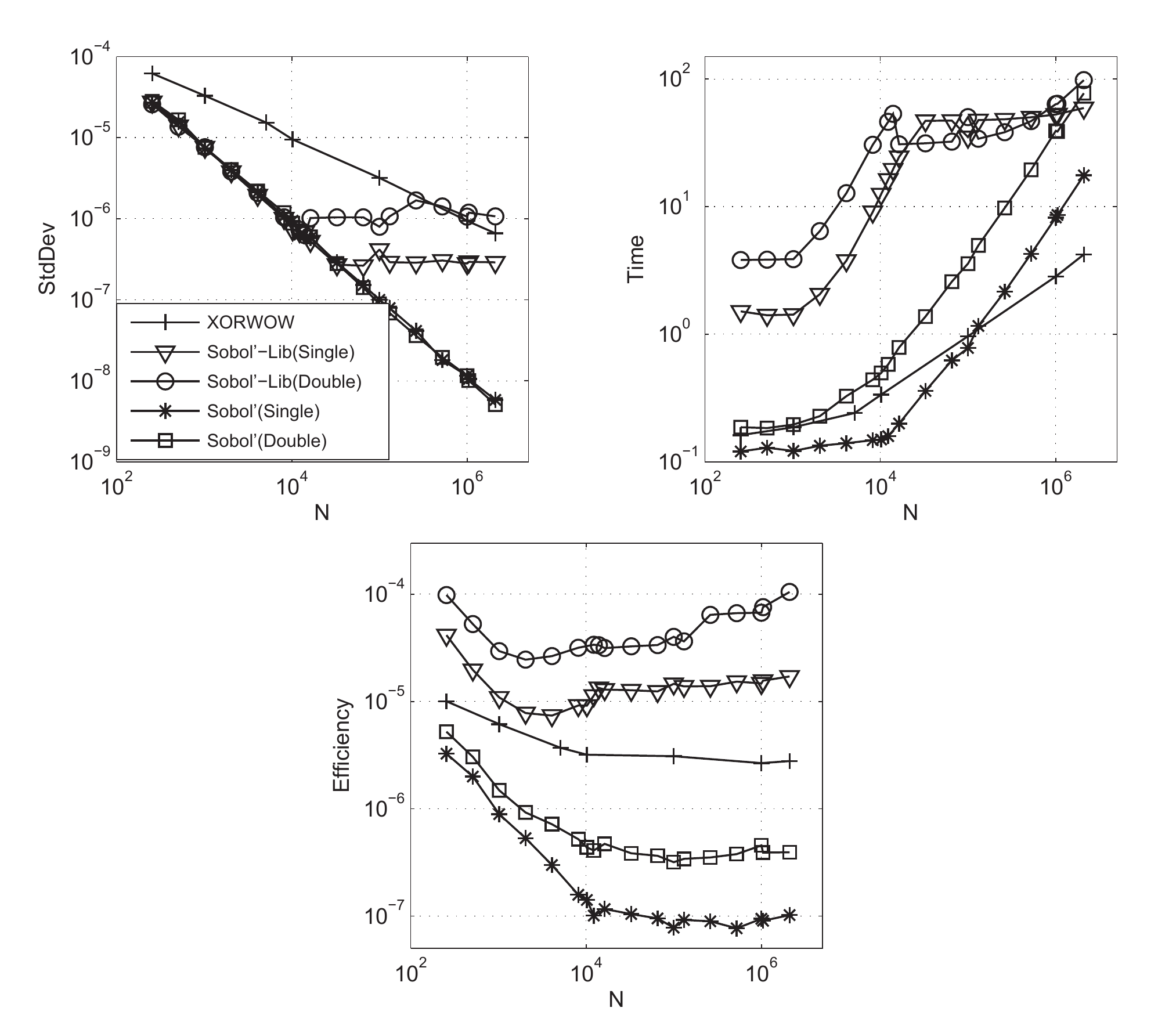}
\end{center}
\caption{Comparing CURAND Sobol' function with our implementation in pricing caplets}
\label{sobolcomp}
\end{figure}

The Sobol' sequence and a scrambled version of it are provided in the CURAND library from NVIDIA. We use both the single precision version (Sobol'-lib(Single)) and double precision version (Sobol'-lib(Double)) in our simulation. We also implement our own version of the Sobol' sequence (Sobol'(Single) and Sobol'(Double)) for comparison. Figure \ref{sobolcomp} plots the sample standard deviation of 100 estimates for the caplet price, computing time, and efficiency, of different implementations of the Sobol' sequence against the sample size $N$. We also include the numerical results obtained using the fast pseudorandom number sequence XORWOW from CURAND as a reference. We make the following observations:

\begin{enumerate}
\item The convergence rate exhibits a strange behavior and levels off for the CURAND Sobol' sequence generators, Sobol'-Lib(Single) and Sobol'-Lib(Double), as $N$ gets large. Our implementation of the Sobol' sequence gives monotonically decreasing sample standard deviation as $N$ increases;

\item The execution time for CURAND generators Sobol'-Lib(Single) and Sobol'-Lib(Double) is significantly longer than our implementation, and not monotonic for a specific range of $N$; 

\item The efficiency of CURAND generators Sobol'-Lib(Single) and Sobol'-Lib(Double) is even worse than the efficiency of the pseudorandom number sequence XORWOW. Our Sobol' sequence implementations have better efficiency than XORWOW.
\end{enumerate}

Due to the poor behavior of the Sobol' sequence in the CURAND library, we will use our implementation of the Sobol' sequence with single precision in the rest of the paper. We will denote this sequence simply as ``Sobol'" in the numerical results.

\subsection{Performance of Rasrap and Sobol' on CPU}
In Section \ref{Section timing comp}, we compared the computing times of several sequences. Here we compare the performance of Mersenne twister, Rasrap and Sobol', when they are used in simulating the LIBOR market model. The sequences are run on one CPU core. 

\begin{figure}

\begin{center}
\includegraphics[
height=4.5in,
width=5.5in
]
{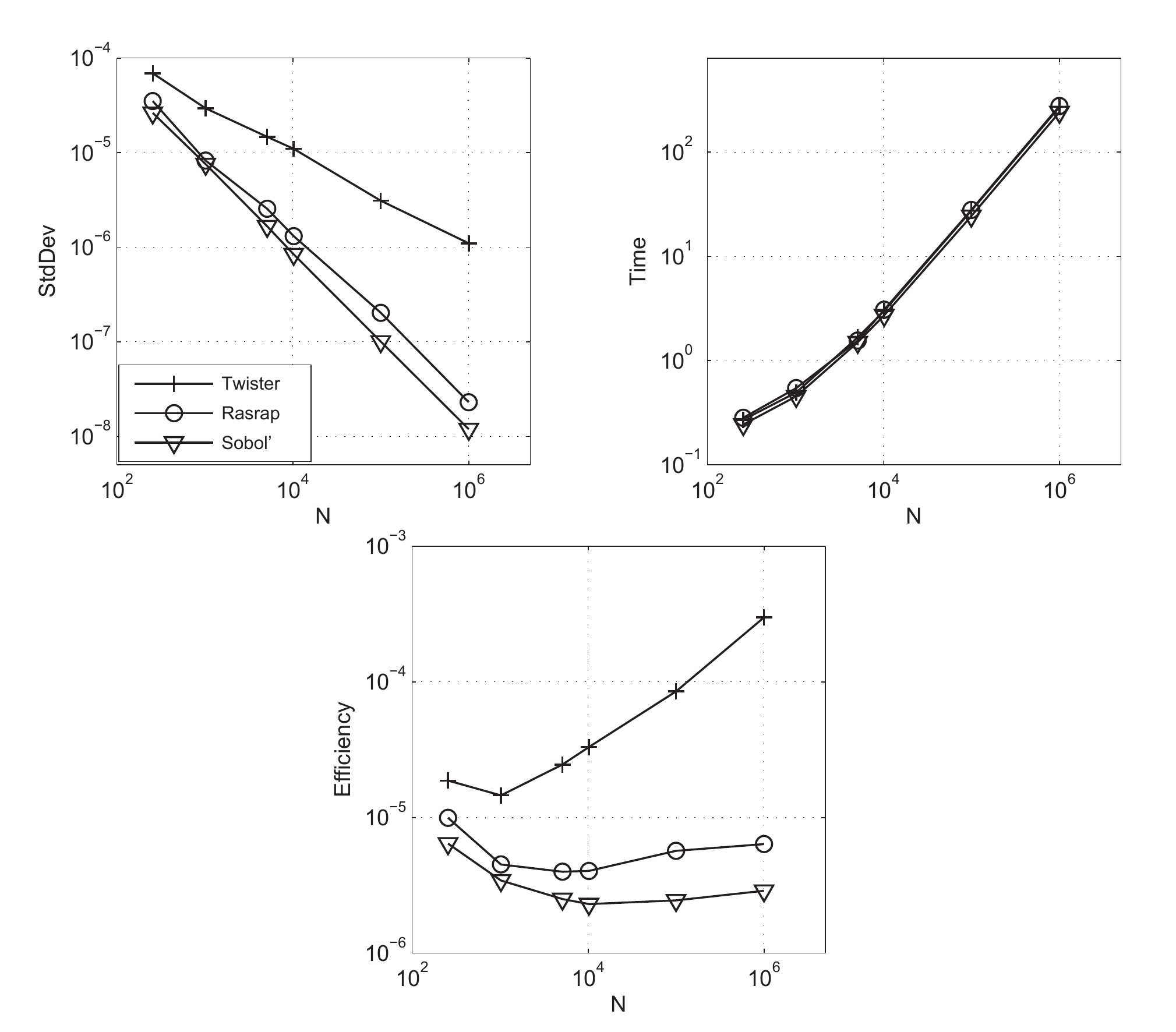}
\end{center}
\caption{Comparing Mersenne twister, Rasrap, and Sobol', when pricing caplets on CPU}
\label{singlecpulibor}
\end{figure}

Figure \ref{singlecpulibor} shows that the sample standard deviation of the estimates obtained from Rasrap and Sobol' sequences converge at a much faster rate than the Mersenne twister. The convergence rate for Mersenne twister is about $O(N^{-0.50})$, and the rate for Rasrap and Sobol' is about $O(N^{-0.87})$ and $O(N^{-0.93})$, respectively. 

The recursive implementation of Rasrap does not introduce much overhead in running time and gives very close timing results to Mersenne twister. The Sobol' sequence based on Gray code is faster than Mersenne twister. As a result, the two low-discrepancy sequences enjoy better and ``flatter" efficiency than that of Mersenne twister.

\begin{figure}

\begin{center}
\includegraphics[
height=4.5in,
width=5.5in
]
{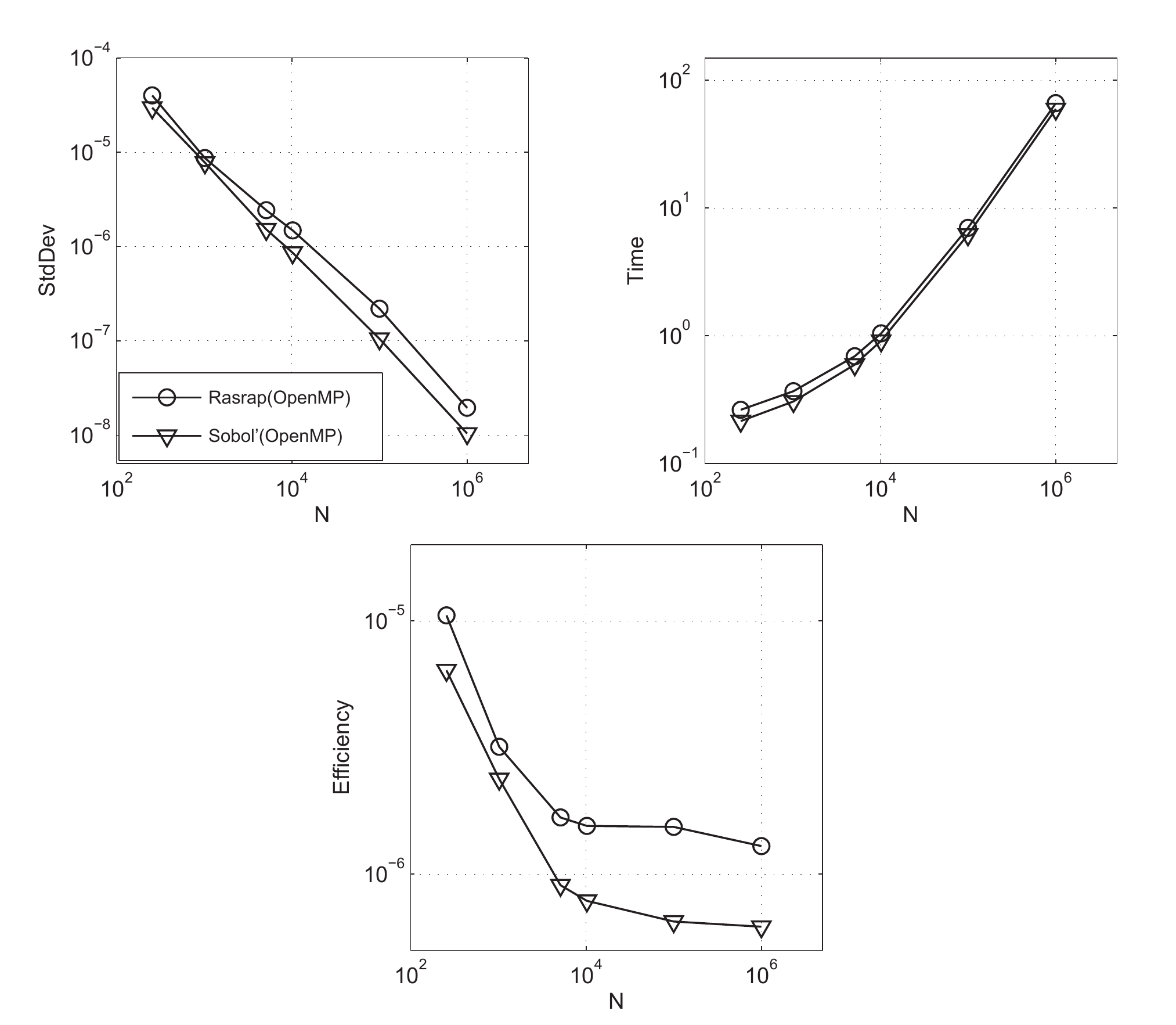}
\end{center}
\caption{Comparing Rasrap and Sobol' when pricing caplets with OpenMP on 8 CPU cores}
\label{liborcpu}
\end{figure}

We next investigate how well Rasrap and Sobol' sequence results scale over multi-core CPU. We implement a parallel version of Rasrap and Sobol' with OpenMP that can run on 8 CPU cores simultaneously. Figure \ref{liborcpu} plots the performance of OpenMP version of Rasrap and Sobol' on CPU. It exhibits the same pattern of convergence, running time, and efficiency as in Figure \ref{singlecpulibor}. The convergence remains the same as in the one core case, but we gain a speedup of four with the parallelism using OpenMP.

\subsection{Performance of GPU}
In this section we compare the counter-based implementations of Rasrap and Sobol' with pseudorandom sequences Philox and XORWOW, on GPU.

\begin{figure}

\begin{center}
\includegraphics[
height=4.5in,
width=5.5in
]
{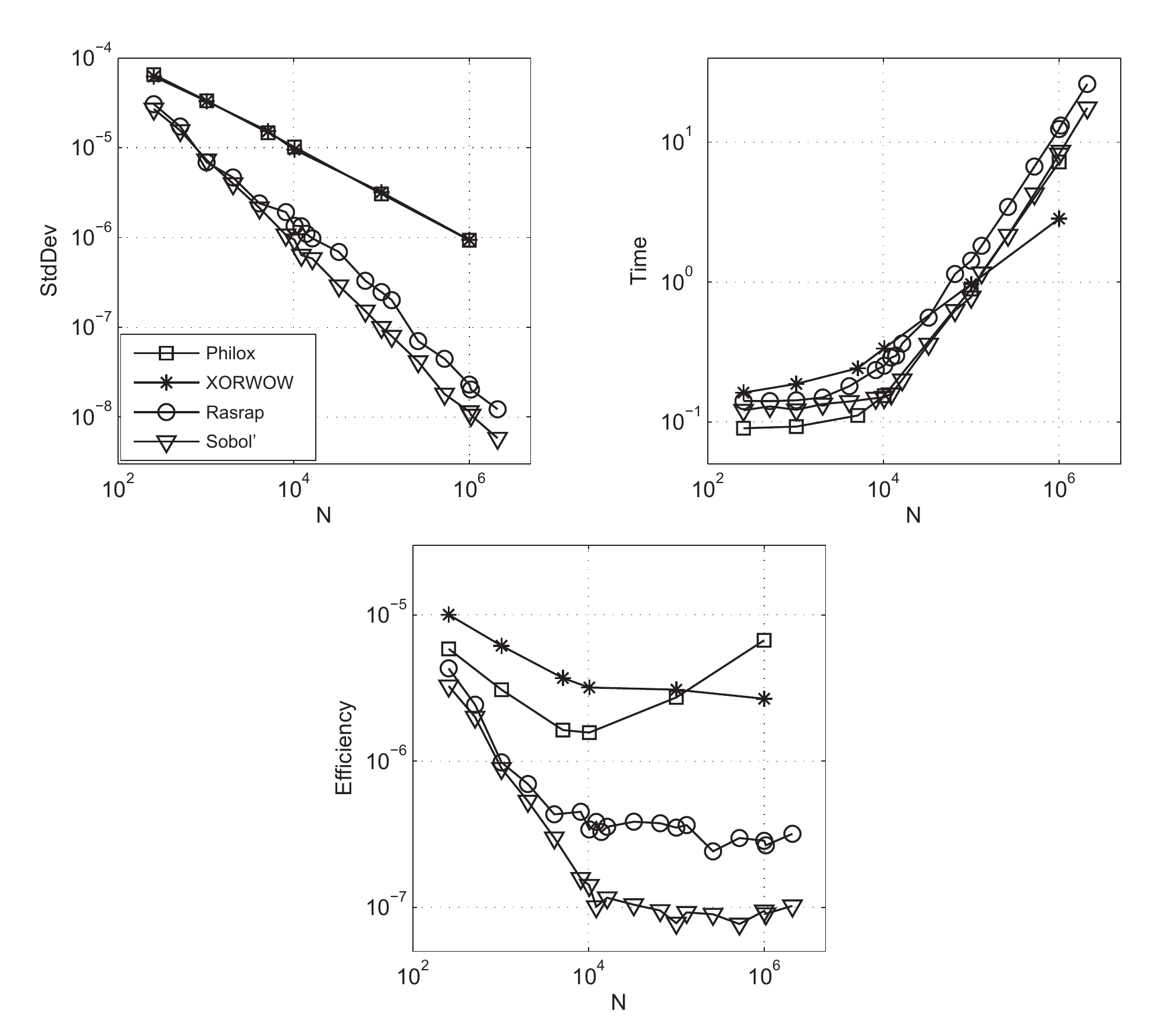}
\end{center}
\caption{Comparing Philox, XORWOW, Rasrap, and Sobol', when pricing caplets on GPU}
\label{liborgpu}
\end{figure}

Figure \ref{liborgpu} plots the sample standard deviation, computing time, and effciency. We make the following observations:

\begin{enumerate}
\item The convergence rate for Philox and XORWOW is about $O(N^{-0.52})$ and $O(N^{-0.51})$, respectively; 

\item The convergence rate for Rasrap and Sobol' is about $O(N^{-0.86})$ and $O(N^{-0.95})$ respectively;

\item XORWOW is the fastest generator, followed by Philox and Sobol'. Rasrap is slightly slower than Sobol';

\item The efficiency of Sobol' is the best among all sequences.

\end{enumerate}

\section{Pricing Mortgage-Backed Securities}
\label{MBS}
We follow the mortgage-backed securities (MBS) model given by \citet{caflisch}. Consider a security backed by mortgages of length $M $ with fixed interest rate $i_{0}$ which is the interest rate at the beginning of the mortgage. The
present value of the security is then

\begin{center}
$PV = E(\mathit{v}) = E(\sum_{k=1}^{M} u_{k}m_{k})$,
\end{center}
where $E $ is the expectation over the random variables involved in the
interest rate fluctuations. The parameters in the model are the following:
\begin{flushleft}
$u_{k} =$ discount factor for month $k $

$m_{k} =$ cash flow for month $k $

$i_{k} =$ interest rate for month $k $

$w_{k} =$ fraction of remaining mortgages prepaying in month $k $

$r_{k} =$ fraction of remaining mortgages at month $k $

$c_{k} =$ (remaining annuity at month $k $) $/c $

$c =$ monthly payment

$\xi_{k} = $$N(0, \sigma)$ random variable.
\end{flushleft}
The model defines several of these variables as follows:

\begin{center}
$u_{k} = \prod_{j=0}^{k-1} (1+i_{j})^{-1} $

$m_{k} = cr_{k}((1 - w_{k}) + w_{k}c_{k})$

$r_{k} = \prod_{j=1}^{k-1} (1 - w_{j}) $

$c_{k} = \sum_{j=0}^{M-k} (1 + i_{0})^{-j} $
\end{center}
The interest rate fluctuations and the prepayment rate are given by

\begin{center}
$i_{k} = K_{0} e^{\xi_{k}} i_{k-1} = K_{0}^{k} e^{\xi_{1}+\cdots+\xi_{k}}%
i_{0}$

$w_{k} = K_{1} + K_{2}\arctan(K_{3}i_{k} + K_{4})$
\end{center}
where $K_{1}, K_{2}, K_{3}, K_{4} $ are constants of the model. The constant
$K_{0} = e^{-\sigma^{2}/2} $ is chosen to normalize the log-normal
distribution so that $E(i_{k}) = i_{0} $. The initial interest rate $i_{0} $
also needs to be specified.

We choose the following parameters in our numerical results:

\begin{center}
$(i_{0}, K_{1}, K_{2}, K_{3}, K_{4}, \sigma^{2}) = (0.007, 0.01, -0.005, 10,
0.5, 0.0004). $
\end{center} 

\begin{figure}

\begin{center}
\includegraphics[
height=4.5in,
width=5.5in
]
{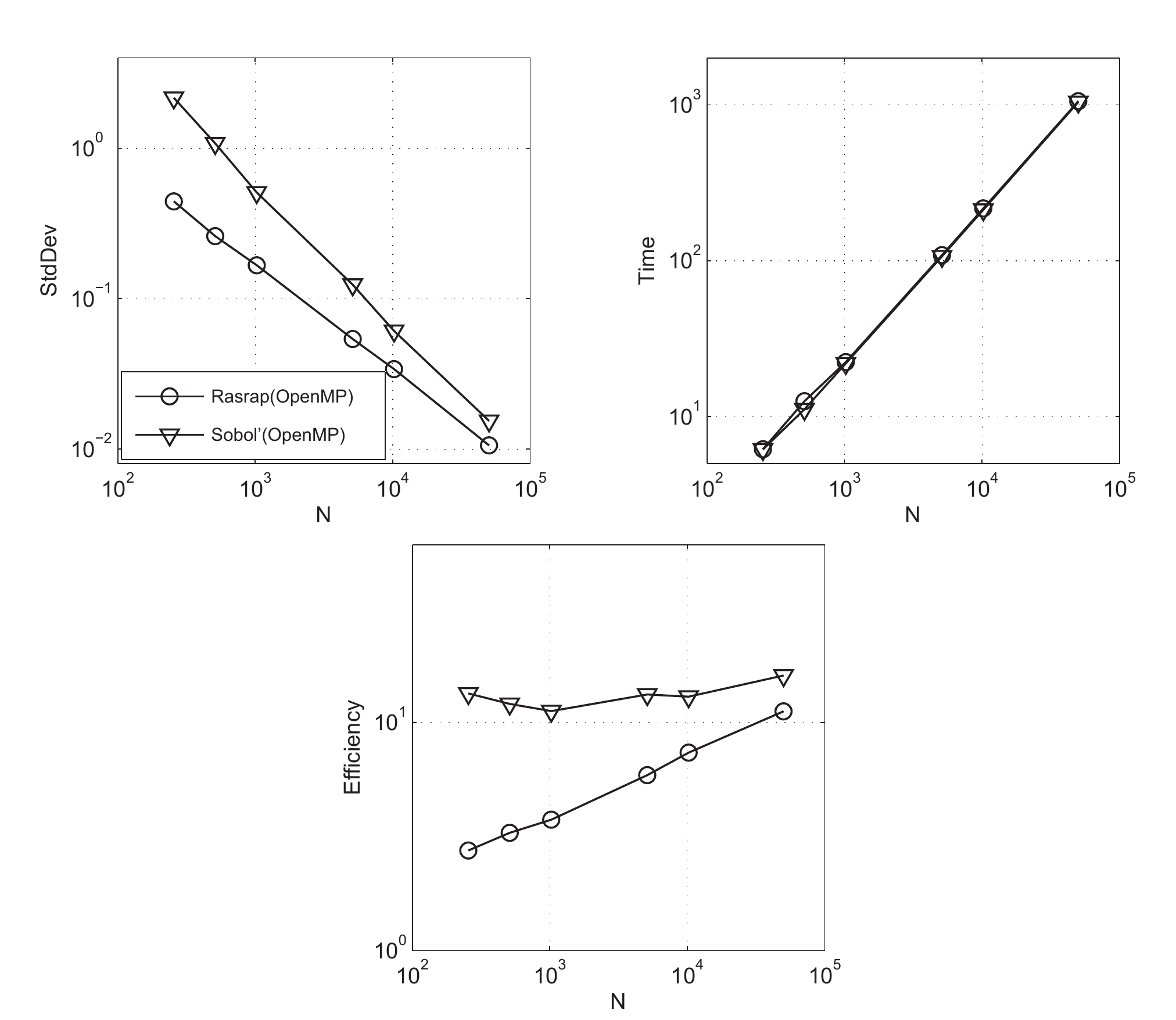}
\end{center}
\caption{Comparing Rasrap and Sobol' when pricing MBS with OpenMP on 8 CPU cores}
\label{mbscpu}
\end{figure}

Figure \ref{mbscpu} compares OpenMP implementations of Rasrap and Sobol' sequences on 8 CPU cores. The sample standard deviation of estimates obtained by Rasrap is smaller than that of Sobol' for every sample size, however, the Sobol' sequence gives a better rate of convergence. We gain a speedup of 6 with the parallelism using OpenMP compared to the single core version. Rasrap has the better efficiency for all sample sizes.

\begin{figure}

\begin{center}
\includegraphics[
height=4.5in,
width=5.5in
]
{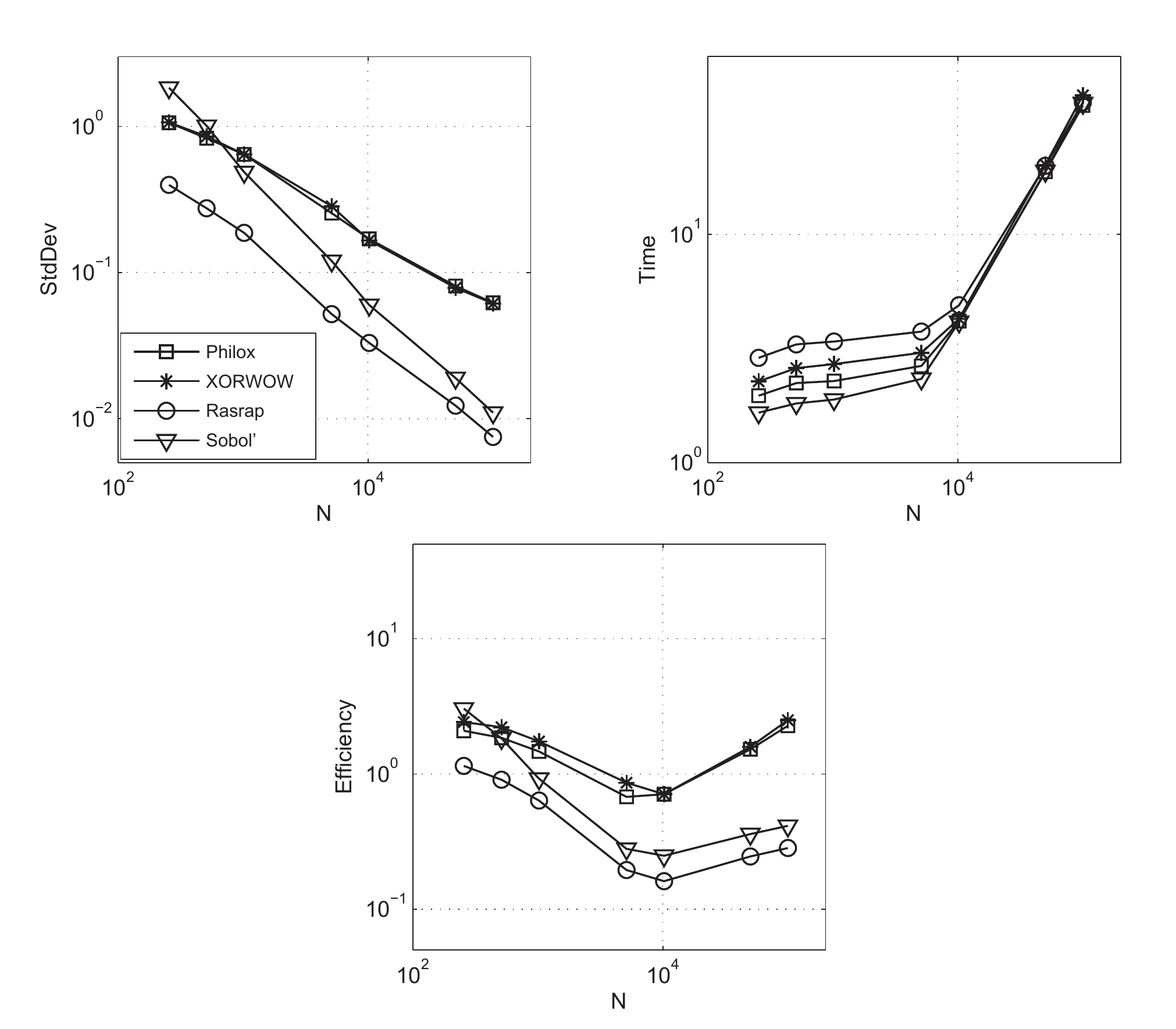}
\end{center}
\caption{Comparing Philox, XORWOW, Rasrap, and Sobol', when pricing MBS on GPU}
\label{mbsgpu}
\end{figure}

Figure \ref{mbsgpu} compares the GPU implementations of Rasrap, Sobol', Philox, and XORWOW. We observe:

\begin{enumerate}
\item The convergence rate for Philox and XORWOW is about $O(N^{-0.5})$;

\item Rasrap gives lower standard deviation than Sobol', however, the convergence rate for Sobol' ($O(N^{-0.86})$) is better than Rasrap ($O(N^{-0.68})$);

\item The efficiency of Rasrap is the best among all sequences.

\end{enumerate}

\section{Comparing GPU and cluster computing }

\begin{figure}
\begin{center}
\includegraphics[width=5.5in, height=4in]{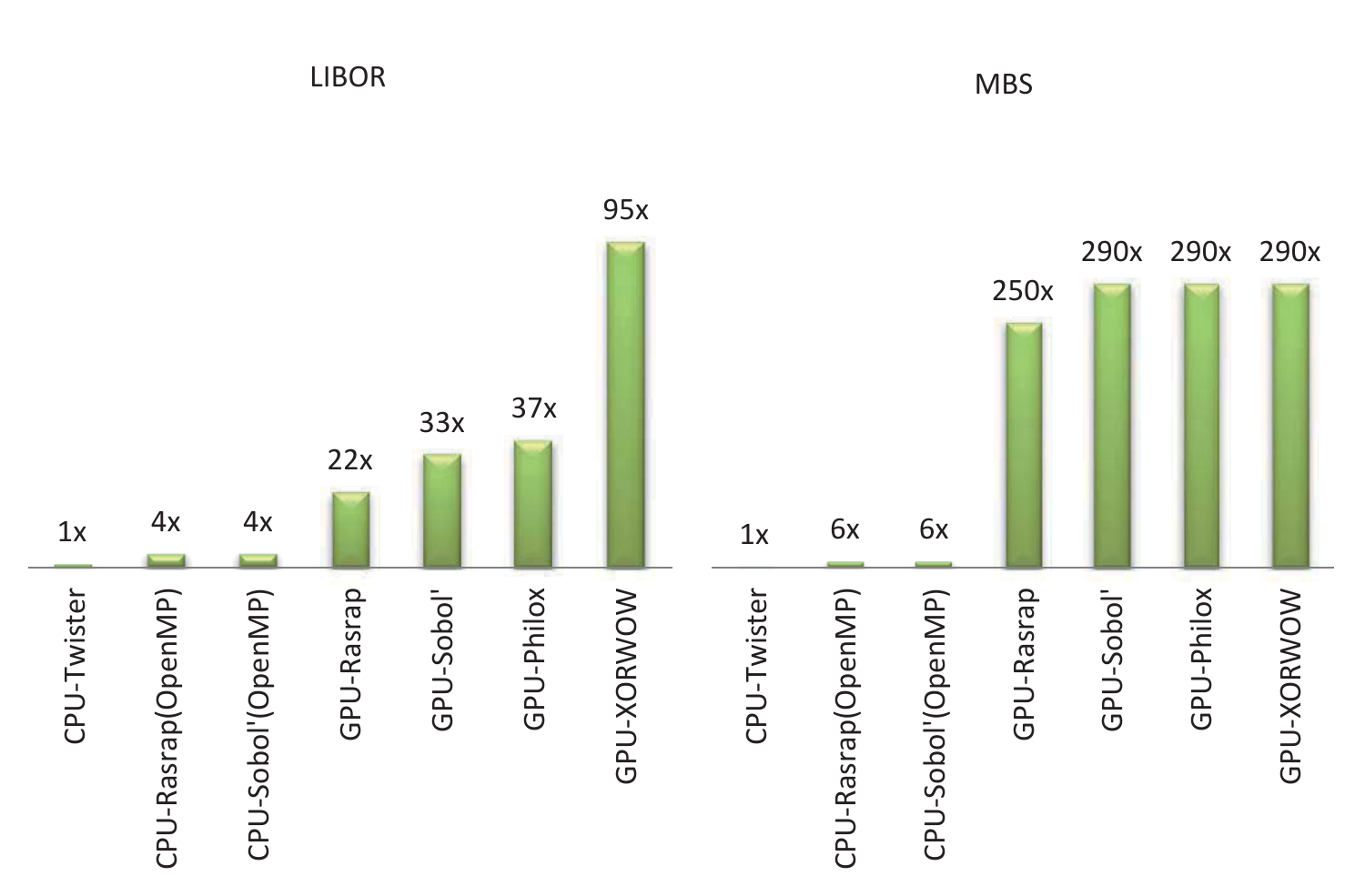}
\end{center}
\caption{GPU speed-up over CPU in pricing LIBOR caplets and MBS}
\label{hist}
\end{figure}

In Figure \ref{hist}, we display the GPU speed-up over CPU for both LIBOR and MBS examples. These results only consider the computing time, and the computing time of CPU-Twister is taken as the base value in each example. The largest speed-up is a factor of 95 and it is due to GPU-XORWOW for the LIBOR market model simulation. In the MBS example, GPU-Rasrap speed-up is a factor of 250, and the other GPU sequences give a speed-up of factor 290.

\begin{figure}

\begin{center}
\includegraphics[
height=3in,
width=5.5in
]
{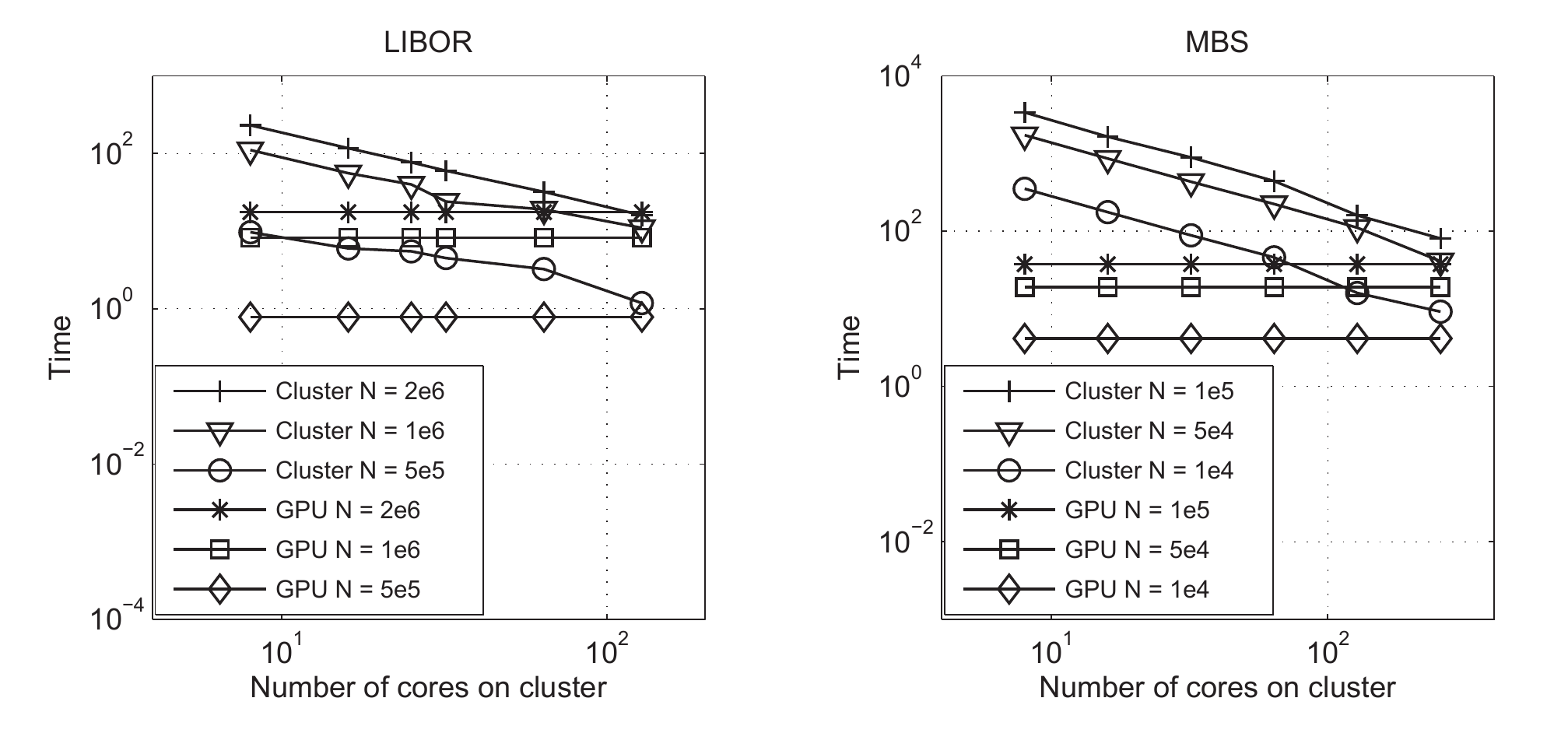}
\end{center}
\caption{Comparing computing time of GPU and FSU HPC cluster in pricing LIBOR caplets and MBS}
\label{cluster}
\end{figure}

Finally, to demonstrate the impressive computing power of GPU, we compare GPU with the high performance computing (HPC) cluster at Florida State University. We implement a parallel Sobol' sequence using MPI, and run simulations for the two examples, LIBOR and MBS. Figure \ref{cluster} plots the computing time against the number of cores used by the cluster, when the sample size $N$ takes various values. The GPU computing time is plotted as a horizontal line since all the cores of GPU are used in computations. Figure \ref{cluster} shows that for the LIBOR example, the GPU we used in our computations has equivalent computing power roughly as 128 nodes on the HPC cluster. This is about when the HPC computing time plot reaches the level of GPU computing time, for each $N$. In the MBS example, 256 nodes on the HPC cluster are equivalent to the GPU. We also point out that on a heterogeneous computing environment such as a cluster, continually increasing the number of nodes will not necessarily decrease the running time due to higher cost of communication between nodes and higher probability that slow nodes are used. But for GPUs, a more powerful product with more cores would suggest gains in computing time.

%%%%%%%%%%%%%%%%%%%%%%%%%%%%%%%%%%%%

\label{lastpage}

\end{document}